\documentstyle[preprint,aps]{revtex}

\def\ni{\noindent}
\def\be{\begin{equation}}
\def\ee{\end{equation}}

\begin{document}

\title {\Large{\bf
Surface Properties of a Selective \\
Dissagregation Model}}

\author{\large Juan R. Sanchez}
\address
{
Dpto. de F\'{\i}sica,
Facultad de Ingenier\'{\i}a, \\
Universidad Nacional de  Mar del
Plata,\\
Av. J.B. Justo 4302, 7600\\
Mar del Plata,
Argentina\\
}


\maketitle

\newpage
\begin{abstract}
The scaling properties of one-dimensional surfaces generated by
selective disaggregation processes are
studied through Monte Carlo simulations.
The model presented here for the deconstruction process takes into
account the nearest neighbors interaction energies.
This interactions are considered to be the energy barrier 
the detaching particle has to overcome in order to desorb 
from the surface.
The resulting one-dimensional surface has self-affine properties and
the scaling exponents of the surface roughness are
functions of the interaction parameter $J/kT$, where $J$ is an
interaction energy, $k$ is the Boltzmann constant and $T$ is the
temperature.
The dependence of the width exponents on the interaction parameter
is analyzed in detail.

\ni PACS: 05.40 68.35Bs 68.35Ct
\end{abstract}

\newpage
There are three fundamental physical processes that gives rise to the
morphology of a surface: deposition, surface diffusion and desorption.
The characteristics of the interfaces generated by the combination of
deposition and surface diffusion
has been well studied during the past decades \cite{fv-b,v-b,bs-b}.
In particular for {\it growth} models, particles are added to the
surface and then are allowed to relax by different mechanisms.
Many of this models have been shown
to lead to the formation of self-affine surfaces, characterized by
scaling exponents.
From a theoretical point of view, the studies dedicated to the
self-affine interfaces generated by growth models
can be considered to follow two main branches.
The studies about the properties of discrete models and the studies
about continuous models.
The first ones where dedicated mainly to the study of the properties of
computational models in which the growth proceeds on an initially empty
lattice representing a d-dimensional substrate.
At each time step, the height of the lattice sites is increased 
by units (usually one unit) representing the incoming particles.
Different models only differs on the relaxation mechanisms proposed 
to capture specific experimental characteristics.
Then, the models are classified according to the values of the scaling 
exponents in several universality classes.
The continuous models, on the other hand, are based on stochastic
differential equations of the type~\cite{vill01}

\be
\partial h/\partial t = F - \nabla .~{\mathbf{j}} + noise \:,
\ee

\ni in which $h(r ,t)$ is the thickness of the film deposited onto the
surface during time $t$, $F$ denotes the deposition rate and
$\mathbf{j}$ denotes the current along the surface which in turn depends
on the local surface configuration. 
When the surface current represents some experimental nonlinear 
equilibrium processes it gives rise to different types of
nonlinear terms.
The {\it noise} term corresponds to the fluctuations in the growth rate 
and, in general, is assumed to be uncorrelated.
The differential equations are solved, either numerically or
analytically (when it
is possible), and the scaling exponents are determined.
If the values of the exponents
are similar to those of the discrete models, it is said that both 
(discrete and continuous models) belong to the same
universality class, although the formal connection between both
approaches is still an open question.

Desorption or detaching processes, despite its technological importance,
have  not received the same attention. 
Maybe because they were considered to be the
reciprocal of the growth models and no new characteristics were
expected to appear.

There are many technological processes depending on the details of the
etching phenomena and
only recently few of them have been modeled in order to understand the
phenomena at an atomic level~\cite{sa01}.
As an example, corrosion is among the most important disaggregation
processes, because its economical importance.
A basic fact is that the corrosion attack is more intense at high
temperatures and strongly depends on the type of material under attack.
It would be desirable to take into account these characteristics in basic
desorption models.
In this paper we present a discrete disaggregation model, which can be
considered as a generalization of a recently reported model~\cite{scalas}.
We claim that our model is much more realistic than previous ones
because we incorporate an interaction parameter proportional to the
nearest-neighbor energies.
The main characteristic of the model presented here is the dependence of
the values of the scaling exponents on the interaction parameter.

It is well known that most chemical reactions that take place at a
surface or interface, as well as those processes that lead to 
particles removing, are activated processes. 
The particle has to overcome an energy barrier in order to
detach. In a first approach, this energy barrier could be consider to be
proportional to the number of bonds the particle has at its location on 
the surface.
Then, the energy barrier could be written as $n J$, where $n$ is the
number of nearest-neighbors of the particle and $J$ is the bonding energy. 
Following this line, it is usual to represent the
probability for a particle to overcome a barrier of height $n J$ as
$p_{(n)} = \exp(-n J/k T)$, where $T$ is the temperature and $k$ is the
Boltzmann constant.
In our model, the resulting self-affine interface shows
different values of the scaling exponents for different values of $J/kT$,
as a consequence of the dependence of the desorption probability on the
interaction parameter.

In our computational model, the surface is represented by an
one-dimensional array of integers specifying the number of atoms 
in each column.
In order to maintain the solid-on-solid characteristic of the model the
detaching processes will not give rise to overhangs.
The simulations were carried out in the following way: each column of the
one-dimensional array is initially filled enough in order to allow 
the deconstruction process to take place during the whole 
simulational time, i.e., no column height must
become negative at the end of the simulation.
By initially filling all columns with the same amount of atoms the 
the simulations start with a flat surface at $t = 0$.
In order to simulate the selectivity of the deconstruction 
process, a column $K$ is picked at random.
Then, the site with the greatest number of open bonds is selected among
the site $s(K)$ lying at the top of the column $K$ and 
the sites $s(K-1)$ and $s(K+1)$
at the top of each of the neighbors columns.
After selecting the column with the greatest number of open bonds, the
detaching probability for the particle at the top of 
the selected column is calculated as $p{(n)} = \exp(- n\:J/kT)$.
The particle is detached from the selected position if a
random number $r$ is less than $p{(n)}$.
Then, the detaching probability depends on the
number of nearest-neighbors $n$ of the atom in the direction $x$
running parallel interface, i.e, in our model $n = 0$, $1$ or $2$.
The interaction energy corresponding to the neigbor in the direction $y$,
perpendicular to the interface, is not taken into account since it is
present in all detaching atoms.
In this way the atom with less neighbors $n$ has the greatest probability to
desorb from the surface.

Following the studies realized to growing models we expect that the
surface
generated by the above described process presents self-affine properties.
Then, we are mainly interested in the
scaling properties of the width (or roughness) defined as

\be
\omega(L,t) = \left[\langle h^2(x,t)\rangle_x - \langle
h(x,t)\rangle_x^2\right]^{1/2}\: ~~,
\ee

\ni where $\langle \cdot \rangle_x$ denotes average over the entire
lattice size
$L$, $h(x,t)$ is the height of the column at position $x$, $\langle
h(x,t)\rangle_x$ is
the average surface height and $t$ is the time, roughly proportional
to the number of removed monolayers $m$.

As for a growth processes we assume that the roughness obeys
the Family-Vicsek scaling law~\cite{fv01}, valid for self-affine interfaces,

\be
\omega(L,t) = L^{\alpha}\:f\left(\frac{t}{L^{z}}\right) ~~,
\ee

\ni being $\alpha$ the roughness exponent because  $\omega \sim
L^\alpha$ when
$t \gg L^{z}$ and $\beta$ the growth exponent
because $\omega \sim t^{\beta}$ when $t \ll L^{z}$. In our case there is,
of course, no growth but an exponent $\beta$ can be defined 
in the same way. $z = \alpha/\beta\:$ is the dynamic exponent.

In our model, the relevant characteristic is the dependence of the
scaling exponents $\alpha$ and $\beta$ on the interaction parameter 
$J/kT$ and this characteristic in investigated in detail here.
The exponent $\beta$ was determined by measuring the slope of the
log-log plot of $\omega(L,t)$ vs $t$ at early times and the results 
presented below were obtained for systems sizes of $L = 5000$.
On the other hand, the exponent $\alpha$ was determined
using the correlation height-height correlation function defined as
$c(r,\tau) = \langle [h(x,t) - h(x+r,
t+\tau)]^2\rangle_{r,\:\tau}^{1/2}$.
The correlation function is known to scale in the same way 
as the width $w(L,t)$; in particular
for $t \gg L^{z}$ and for $r \ll L$ the correlation function behaves as
$c(r,0) \sim r^\alpha \:$.
Results presented below for the exponent $\alpha$ where
obtained by measuring the slope of the log-log plot of 
$c(r, \tau)$ vs $r$ in the limit $r \rightarrow 0$.
For all the presented values of the interaction parameter $J/kT$ the
used lattice size was $L = 1000$ because the correlation function
calculation must be done at long times in order to allow the width to
reach the saturation value.
Averages where taken over the necessary independent runs to reduce the
statistical error below $0.4 \%$ in all the simulations.

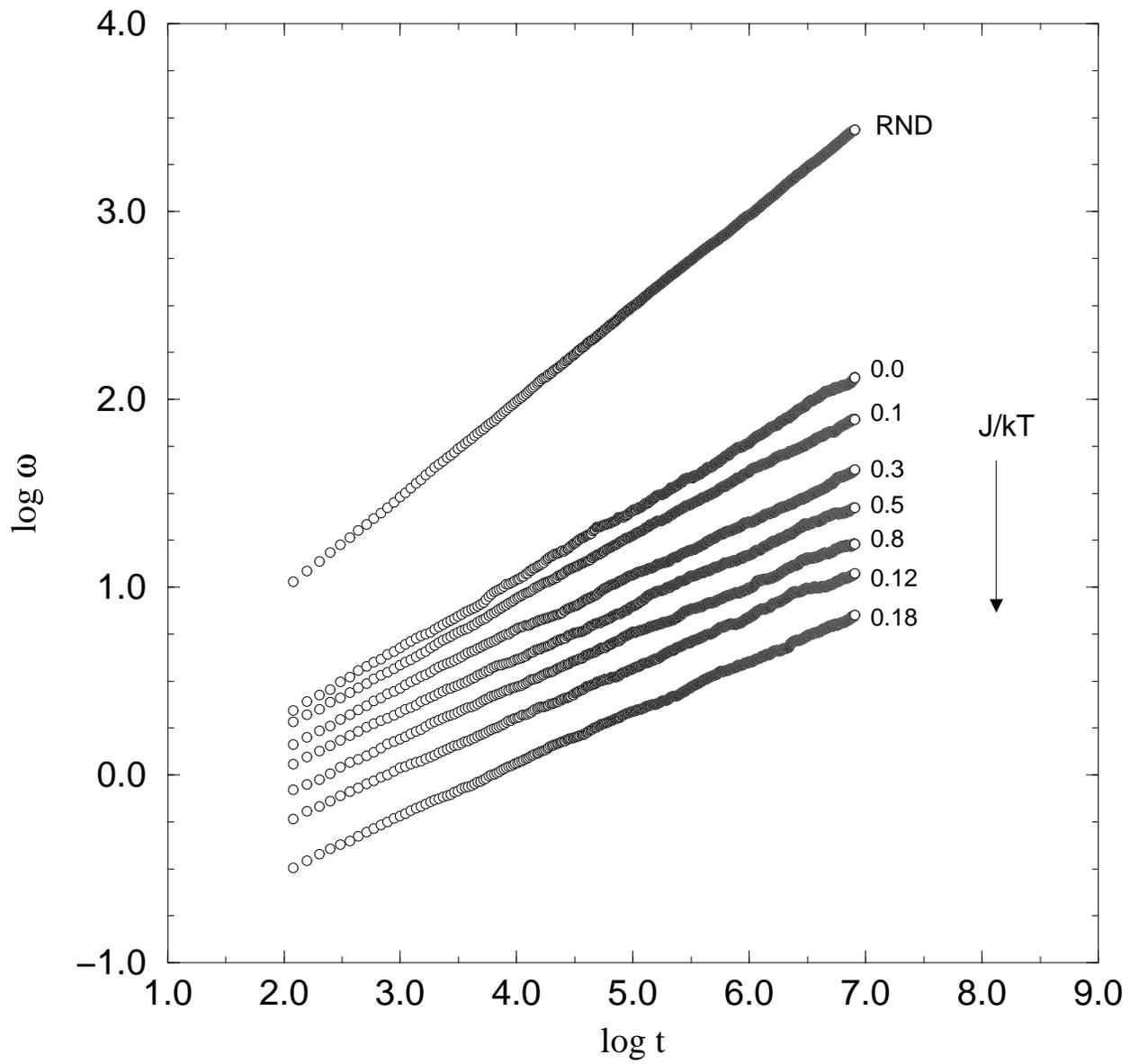
\begin{figure}                                                                                                  
\caption{Plot of the interface width $w(L,t)$ versus time in  log-log scale.                                              
The slope through the upper points with the label RND is 0.5. The other points
correspond to the simulation results of the deconstruction model, for various
values of $J/kT$, using $L=5000$.}                                                                                                      
\label{fig1}                                                                                                 
\end{figure}

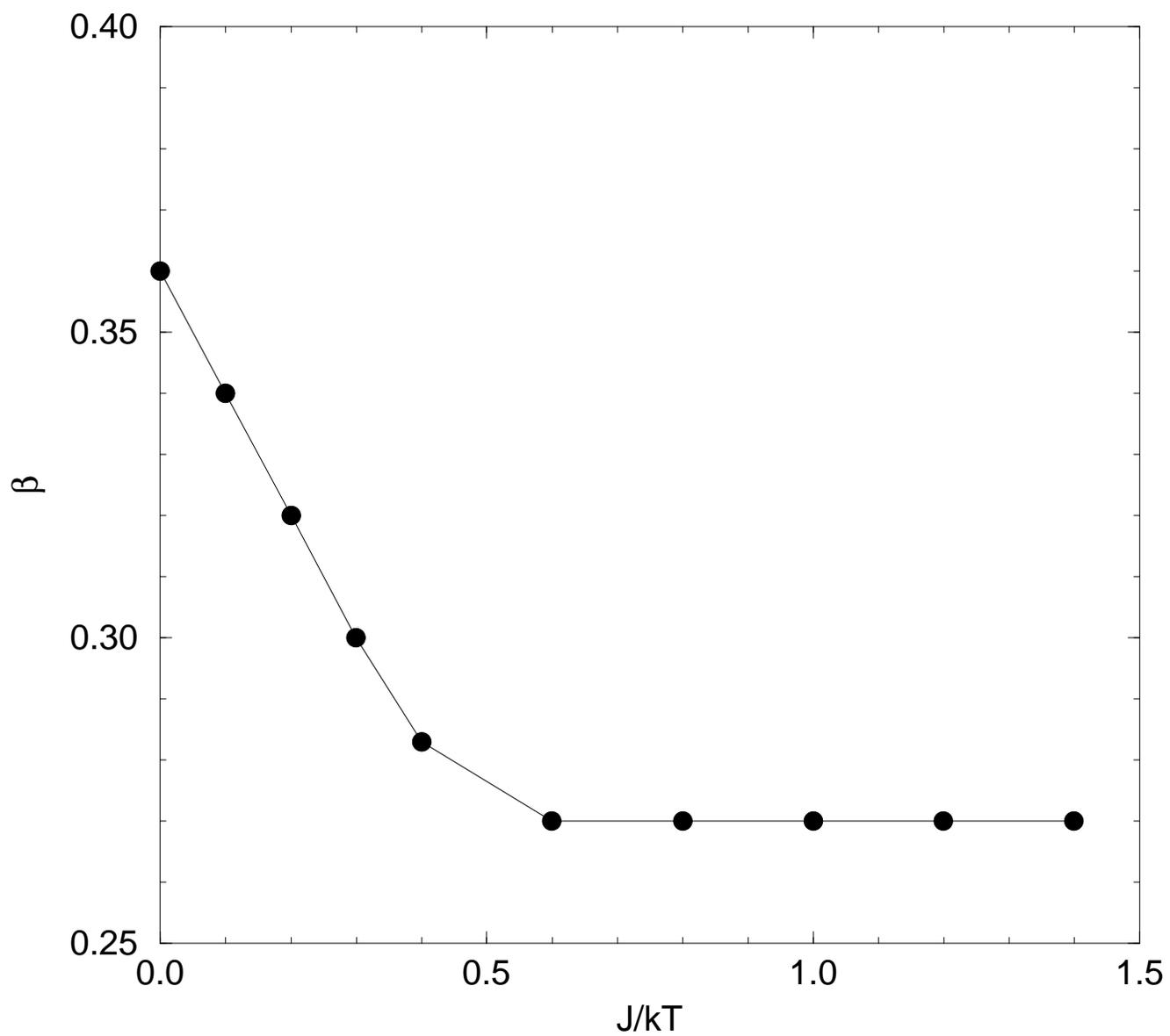
\begin{figure}                                                                                                  
\caption{The exponent $\beta$, measured from the slopes through the points
plotted in Fig.~\ref{fig1}, as a function of $J/kT$.}                                                                                                      
\label{fig2}                                                                                                 
\end{figure}

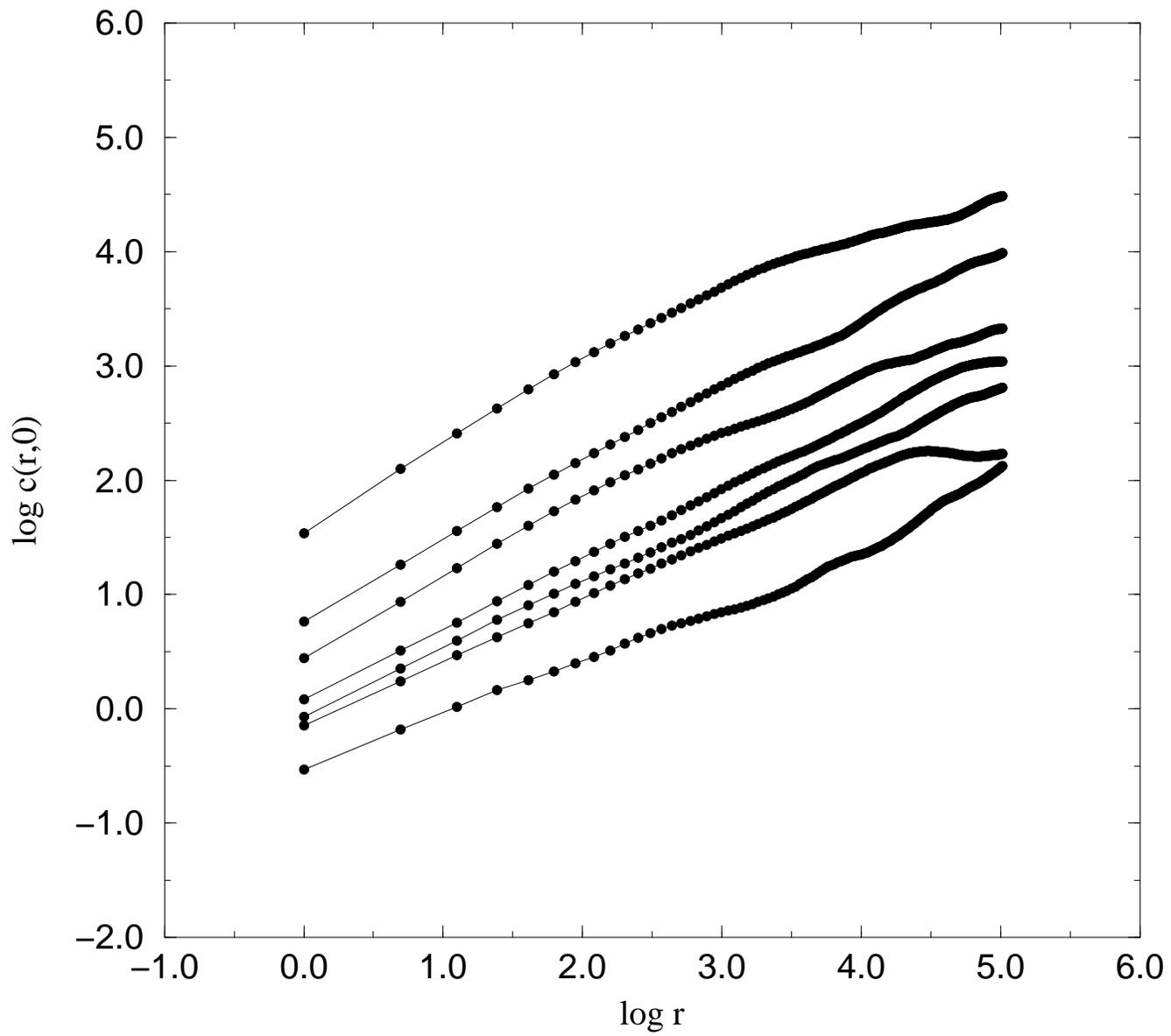
\begin{figure}                                                                                                  
\caption{Height-height correlation function $c(r,0)$ as a function of $r$
for various values of $J/kT$ plotted in a log-log scale.
The correlation function is measured using $L=1000$ after the saturation
value of the width was reached.}                                                                                                      
\label{fig3}                                                                                                 
\end{figure}

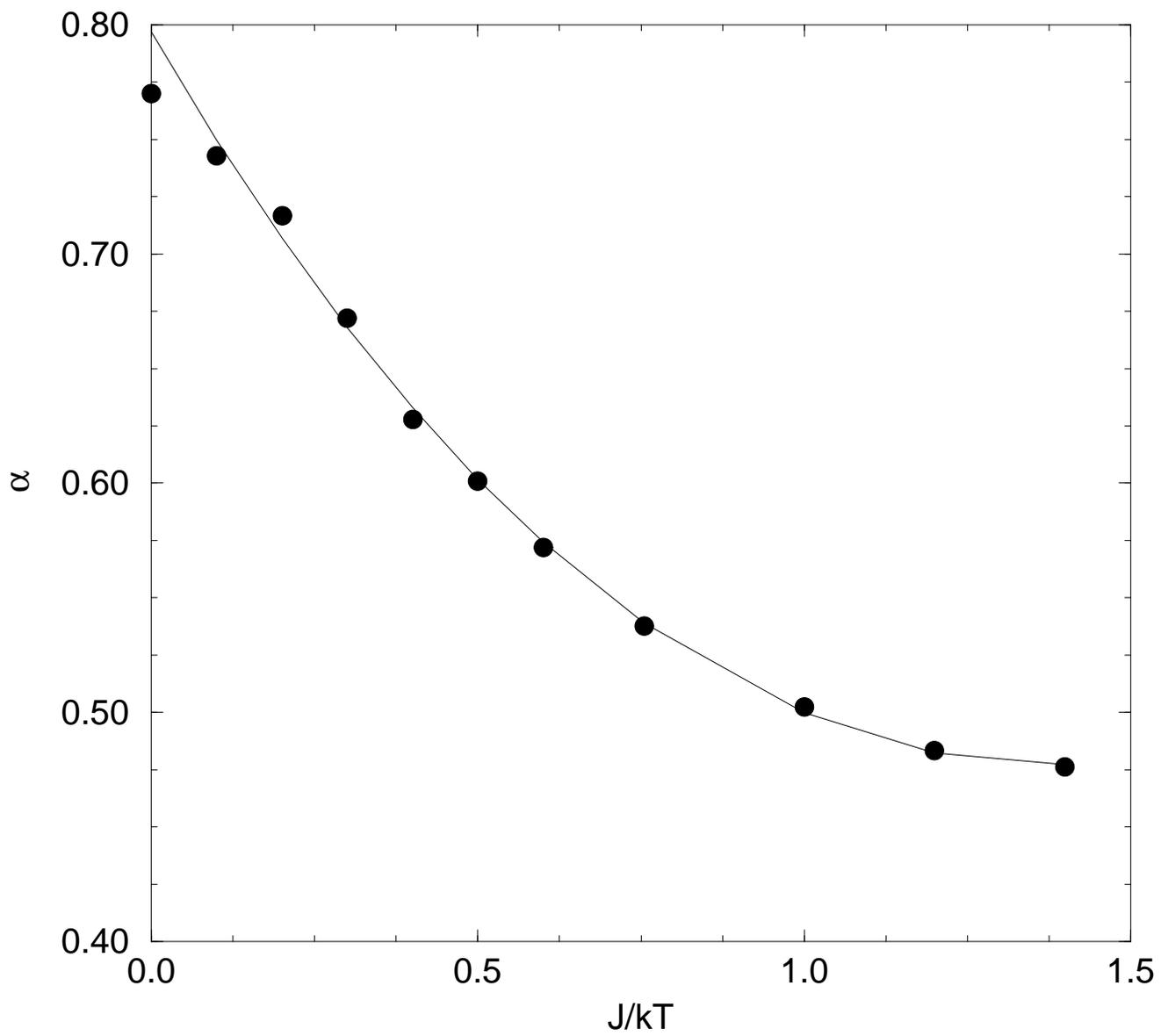
\begin{figure}                                                                                                  
\caption{The exponent $\alpha$, measured from the slopes through the points
plotted in Fig.~\ref{fig3} in the $r \rightarrow 0$ limit,
as a function of $J/kT$.}
\label{fig4}                                                                                                 
\end{figure}

In Fig.~\ref{fig1} a log-log plot of the interface width versus time is presented.
Time is advanced by $1/L$ when a particle at the upper monolayer 
is asked to detach.
In Fig.~\ref{fig1} the upper points with the label RND where obtained by removing the
particle at the top of randomly selected columns.
Its slope is 0.5 and is presented as a reference.
The other plots corresponds to the behavior of the width for various
increasing values of the interaction parameter $J/kT$.
It can be seen  that the slope is smaller for larger values of $J/kT$.
This behavior is quantitatively verified by the plot in Fig.~\ref{fig2}. 
In this graphic, the values
of the exponents $\beta$ are plotted as a function of $J/kT$. 
The values for $\beta$ were
obtained from the slopes of the lines fitting the log-log plot of the
points of the width vs time. 
Although only few values of $J/kT$ are plotted, clearly there is a
continuous variation of the exponent $\beta$ with the interaction parameter.
It can be seen that the value of $\beta$ deceases as the interaction
parameter increases but, after certain crossover, a limiting value 
$\beta \cong 0.26$ in reached and the exponent does
not decrease for further increments in $J/kT$.
The height-height correlation function for long times is
plotted in Fig.~\ref{fig3} for the same values of $J/kT$ 
as shown in Fig.~\ref{fig1}. 
The dependence of the slope of $c(r,0)$ on the interaction
parameter is evident. 
The values of the exponent $\alpha$ were obtained
from the slope of the lines through the points for $r \rightarrow 0$. 
The values of $\alpha$ are plotted in Fig.~\ref{fig4} as a function of $J/kT$. 
As for $\beta$, the same limiting effect can be observed, and also the value
of the exponent decreases as the interaction increases for the lowest 
values of the parameter.

The presented results can be interpreted in several ways. 
For fixed $J$, these results show that below certain temperature 
the scaling properties of the interface does not change. 
For fixed temperature it can be seen that materials with greater interaction
energies present more resistance to etching. 
However, these are not linear effects.
On the other hand, the behavior of the width of the surface is in 
agreement with the intuitive picture of a corrosion process, for instance. 
The roughness of the surface tends to decrease
with increments of the interaction parameter indicating that
if the interaction energy is large or the temperature is low, the surface 
is less affected by the attack, but eventualy exponent $\beta$ reaches a saturation
value. 

In summary, the model presented here capture the basic behavior of
chemical disaggregation processes and show how the self-affine 
characteristics of the surface are
affected by the inclusion of an interaction parameter.
The model cannot be associated to any known continuous
model because there is a parameter that continuously control 
the value of the scaling exponents.
No continuous model shows this characteristic.
In continuous {\it growth} models the values  of the
exponents are independent of any parameter involved in the stochastic
differential equations.
A possible equivalent discrete growth model would be a deposition model in
which the particles sticks at the sites with the smaller number of bonds.
However this will result in an unrealistic deposition model.
The model presented here incorporates a realistic physical parameter, the
interaction $J/kT$, which appears usually in many atomistic processes and, 
since the values of the scaling exponents depend on this parameter, the model 
does not fall in any known universality class.
There are interested extensions of the present work that can be pointed out; 
the inclusion of next-nearest-neighbors interactions (competitive and not competitive) 
and the case of deconstruction in $2+1$ dimensions. These cases are 
under study and the results will be published elsewhere.

\end{document}